\documentclass[aps,pra,reprint,superscriptaddress,
 longbibliography,floatfix]{revtex4-2}

\usepackage{hyperref}
\hypersetup{colorlinks=true, citecolor=blue, urlcolor=blue, linkcolor=blue}
\usepackage{graphicx}
\usepackage{bm,amsmath}

\usepackage[dvipsnames]{xcolor}

\begin{document}

\title{Dipole modes of a trapped bosonic mixture: Fate of the sum-rule approach}

\author{Chiara Fort} 
\affiliation{\mbox{Dipartimento di Fisica e Astronomia and LENS, Universit\`{a} di Firenze, 50019 Sesto Fiorentino, Italy}}
\affiliation{Istituto Nazionale di Ottica, CNR-INO, 50019 Sesto Fiorentino, Italy}

\author{Michele Modugno}
\affiliation{Department of Physics, University of the Basque Country UPV/EHU, 48080 Bilbao, Spain}
\affiliation{IKERBASQUE, Basque Foundation for Science, 48013 Bilbao, Spain}
\affiliation{EHU Quantum Center, University of the Basque Country UPV/EHU, 48940 Leioa, Biscay, Spain}

\date{\today}

\begin{abstract}
We present a general discussion of the dipole modes of a heteronuclear bosonic mixture in a harmonic trap, by comparing the prediction of the sum-rule approach with full Gross-Pitaevskii (GP) calculations. 
Yet in the range of interaction in which the mixture is miscible and stable at the mean-field level, $g_{12}^2<g_{11}g_{22}$, we find that there are regimes where the sum rules display significant deviations from the GP solution. 
We provide an interpretation for this behavior in terms of the Ehrenfest approach, which we show to be equivalent to the sum-rule approach for dipole excitations.
\end{abstract}

\maketitle

\section{Introduction}

Sum rules are a powerful and widely employed tool for investigating the collective modes in different many-body systems, including nuclear collective states \cite{leonardi1971,bohigas1979} and giant resonances in nuclei \cite{lipparini1989}, Fermi liquids at low temperatures \cite{stringari1984}, free electrons in metal spheres, and cavities and shells \cite{lipparini1996}, among others. 
Since the first experimental realization of Bose-Einstein condensation in trapped dilute gases (see, e.g., \cite{dalfovo1999}), and the subsequent explosion of ultracold atom physics, the use of the sum-rule approach has become customary in this field \cite{stringari1996,vichi1999,miyakawa2000,modugno2003,modugno2006,maruyama2006,banerjee2007,maruyama2008,banerjee2009,li2012,ferrier-barbut2014,sartori2015,bienaime2016,FerrierBarbut2018,parisi2018}. 
In particular, it has been broadly used for describing collective excitations of Bose-Fermi \cite{miyakawa2000,maruyama2006,banerjee2007,maruyama2008,banerjee2009,ferrier-barbut2014} and Bose-Bose \cite{li2012,sartori2015,bienaime2016} mixtures.
The advantage of this approach is that the frequency of the collective modes can be estimated by means of algebraic techniques, without having to solve the equation of motion, based only on the knowledge of the ground state solution \cite{pitaevskii2016}. 
At the same time, this also constitutes a limit of this approximate method, since it cannot account for dynamical effects that may break the self-similarity of the density distribution.

In order to shed some light on the above point, as a proof of concept here we consider the case of the dipole mode of a mixture of Bose-Einstein condensates (BECs) that was recently considered in a number of experiments \cite{zhang2012,bienaime2016,wilson2021,cavicchioli2022}. 
It is worth noticing that the collective excitation spectrum of a two-species BEC has been originally addressed within the framework of the Bogoliubov theory \cite{pu1998}, but a general discussion of the sum-rule approach for this case is still lacking, contrary to the case of Bose-Fermi mixtures. 
The latter was first considered in Ref. \cite{miyakawa2000} for the case of symmetric mixtures (same mass and trapping frequency) with tunable interactions, and then extended to the heteronuclear case (different masses) in Ref. \cite{banerjee2009}. 
Remarkably, in Refs. \cite{maruyama2006,maruyama2008} it was also shown that the sum-rule approach may not always provide an accurate prediction for the dipole oscillation frequencies as obtained from the full dynamical equations. 
This deviation is particularly noticeable in the repulsive case, especially for the fermionic component. 
Indeed, the different statistics of the Bose-Fermi mixture and the large spatial extension of the Fermi clouds resulting in different regions where the two clouds are superimposed or not, further complicates the coupled dipole dynamics as the Fermi component may be characterized by several oscillations modes \cite{maruyama2008}.

In light of the above considerations, it is interesting to address to what extent the sum-rule approach can be considered a suitable approximate method for describing the collective dipole oscillations of a bosonic mixture. 
To this end, we first extend the sum-rule formalism of Refs. \cite{miyakawa2000,banerjee2009} to the general case of a Bose-Bose degenerate gas mixture with different masses, frequencies, and atom numbers. 
Then, we show that the sum-rule predictions for the dipole frequencies can also be obtained from an effective approach based on the Ehrenfest theorem \cite{ehrenfest1927}, which also provides the amplitudes of the modes and turns out to be helpful also for understanding the limits of validity of these approximations. 
These results are compared with the frequency and amplitudes of the dipole modes extracted from the full Gross-Pitaevskii dynamics as a function of the interspecies interaction, in the whole regime in which the mixture is miscible and stable against mean-field collapse. 

Motivated by a recent experiment by Cavicchioli \textit{et~al.} \cite{cavicchioli2022}, we consider in particular the case of a $^{41}$K-$^{87}$Rb mixture, discussing also the effect of the atom number imbalance. 
We find that, for a balanced mixture, the sum-rule and Ehrenfest approaches provide an accurate prediction for the low-lying dipole mode, from the strongly attractive regime up to weakly repulsive interactions. 
Instead, the high-lying mode is well reproduced in the opposite regime, from weak attraction up to the repulsive regime, provided the system does not enter (local) phase separation. 
This sort of specularity stems from the fact that the two oscillation modes exchange their character from in-phase to out-of-phase behavior when crossing the noninteracting point at $a_{12}=0$. 
For imbalanced mixtures, we find that the accuracy of the sum-rule and Ehrenfest approaches gets reduced depending on the degree of superposition of the two clouds, similar to the case of a Bose-Fermi mixture \cite{maruyama2008}. 

This paper is organized as follows. In Sec.~\ref{sec:general} we present the general formalism, namely we review the sum-rule approach (Sec.~\ref{sec:sumrules}), we introduce the GP formalism, and we discuss the effective description of the dipole dynamics based on the Ehrenfest approach (Sec.~\ref{sec:Ehrenfest}). 
In Sec.~\ref{sec:results} we work out a thorough comparison between the exact GP results and the predictions of the sum-rule and Ehrenfest approaches, for the case of a heteronuclear Rb-K mixture. Final remarks are given in the conclusions.

\section{General formalism}
\label{sec:general}
Let us consider a binary mixture of degenerate gases,
each consisting of $N_i$ atoms with masses $m_i$ ($i=1,2$),
confined in harmonic potentials $V_{i}^{ho}$ with frequencies $\omega_{ki}$ ($k=x,y,z$). The total Hamiltonian of the system can be written as
\begin{equation}
H=\sum_{i=1}^2 H_{i0} + H_{int},
\end{equation}
with
\begin{equation}
H_{i0}=\sum_{\ell=1}^{N_i}\left[
 \frac{\bm{p}_{i\ell}^2}{2m_i} + V_{i}^{ho}(\bm{r}_{i\ell})
+g_{ii}\sum_{\ell'<\ell}\delta(\bm{r}_{i\ell}-\bm{r}_{i\ell'})\right],
\end{equation}
\begin{equation}
H_{int}=g_{12}\sum_{\ell=1}^{N_1}\sum_{\ell'=1}^{N_2}
\delta(\bm{r}_{1\ell}-\bm{r}_{2\ell'}).
\end{equation}
The coupling constants are $g_{ii}=4\pi\hbar^2a_{ii}/m_i$ 
and $g_{12}=2\pi\hbar^2a_{12}/\mu$, with $\mu=m_1m_2/(m_1+m_2)$
being the reduced mass of the system and $a_{12}$ being the interparticle scattering lengths.

The dipole operator along a certain direction, let us say $x$, can be defined as 
\begin{equation}
D[\alpha_1,\alpha_2]\equiv\sum_{i=1}^2\alpha_i\sum_{\ell=1}^{N_i}x_{i\ell}
\equiv\sum_{i=1}^2\alpha_i D_i,
\end{equation}
where $\alpha_i$ parametrize the mixing between the two species.

\subsection{Sum-rule approach}
\label{sec:sumrules}

An estimate of the oscillation frequencies corresponding to the above dipole operator
can be obtained by means of the sum-rule approach, in terms of the energy weighted moments $\mu_1$ and $\mu_3$
\cite{bohigas1979,stringari1984,lipparini1989,miyakawa2000},
\begin{equation}
\omega^2=\frac{1}{\hbar^2}\frac{\mu_3}{\mu_1},
\end{equation}
which can be evaluated in terms of the commutators
\begin{eqnarray}
\mu_1&=&\frac{1}{2}\langle\left[D,[H,D]\right]\rangle,\\
\mu_3&=&\frac{1}{2}\langle\left[[D,H,],[H,[H,D]]\right]\rangle,
\end{eqnarray}
yielding
\begin{equation}
\omega^2=\frac{\displaystyle\sum_{i=1}^2\frac{\alpha^2_i N_i}{m_i} 
\left(\omega^2_{i}-\frac{g_{12}I}{m_i N_{i}}\right)
+2\alpha_1\alpha_2\frac{g_{12}I}{m_1m_2}}{
\sum_{i=1}^2{\alpha^2_i N_i}/{m_i}},
\label{eq:omega_alpha}
\end{equation}
where $\omega_{i}\equiv\omega_{xi}$ (the result is independent of the trapping frequencies along the directions orthogonal to that of the dipole operator $D$) and the volume integral 
$I\equiv\int\partial_{x}n_{01}\partial_{x}n_{02}d^3r$ is determined by the ground-state configuration of the system characterized by the densities $n_{0i}$. 
This clearly highlights the pros and cons of the sum-rule approach: On the one hand, it provides an estimate of the dipole mode frequencies on the basis of only \textit{static} properties; on the other hand, it may fail if the system configuration substantially changes during the \textit{dynamical} evolution (see below).

Following Ref.~\cite{miyakawa2000} the weights $\alpha_i$ can be conveniently parametrized as 
$\alpha_1=\left(\cos\theta+\sin\theta\right)/\sqrt{2}$, 
$\quad\alpha_2=\left(\cos\theta-\sin\theta\right)\sqrt{2}$, so that the total dipole operator $D$
can be written as a linear combination of in-phase and
out-of-phase oscillations of the two species, with $\theta$ playing the role of the mixing angle 
\begin{equation}
D=\frac{1}{\sqrt{2}}(D_1+D_2)\cos\theta
+\frac{1}{\sqrt{2}}(D_1-D_2)\sin\theta.
\label{eq:mixing}
\end{equation}
Then, Eq.~(\ref{eq:omega_alpha}) becomes
\begin{equation}
\omega^2(\theta)=\frac{B_+ +B_-\sin2\theta +2\gamma_{12}\cos2\theta}{
A_+ + A_-\sin2\theta},
\label{eq:omega}
\end{equation}
with $\gamma_{12}={g_{12}I}/{m_1m_2}$ and
\begin{align}
A_\pm&=\frac{N_1}{m_1}\pm\frac{N_2}{m_2},
\label{eq:a-term}\\
B_\pm&=\left(\frac{\omega_1^2N_1}{m_1}\pm\frac{\omega_2^2N_2}{m_2}
\right)-g_{12}I\left(\frac{1}{m_1^2}\pm\frac{1}{m_2^2}
\right).
\label{eq:b-term}
\end{align}
The above set of equations (\ref{eq:omega})-(\ref{eq:b-term}) constitutes the analog of the Bose-Fermi results of Refs.~\cite{miyakawa2000,banerjee2009}, here for the case of a bosonic mixture trapped in a generic harmonic potential (with the two species having different masses, frequencies, and atom numbers)
\footnote{One can easily check that in the special case of spherically symmetric potentials,
with the same frequencies ($\omega_1=\omega_2\equiv\omega_0$) and masses ($m_1=m_2\equiv m$) for the two species, Eq.~(\ref{eq:omega}) reproduces the result in Eq.~(6) of Ref. \cite{miyakawa2000}, namely $\omega^2(\theta)=\omega_0^2-({4g_{12}I}/{m})\sin^2\theta/[
N_1 + N_2 + (N_1 - N_2)\sin2\theta]$, with $I=\Omega/3$}.

Finally, 
one has to choose a criterion for determining the possible mixing angles and the corresponding frequencies. First of all, since we have two components, two different modes are expected, one ``in phase'' and the other one ``out of phase'' (see also the discussion in the next section). Then, considering that sum rules provide an upper bound for the mode frequencies (see, e.g., Ref. \cite{pitaevskii2016}), it is natural to identify the mixing angle of the low-lying mode with the one that minimizes $\omega^2(\theta)$, for which $\partial\omega^2/\partial\theta=0$. In Ref. \cite{miyakawa2000} the other mode was taken to be orthogonal to the former, but this is not necessarily so, because in general they are eigenmodes of a non-symmetric, but diagonalizable, matrix. This will be clear from the discussion in  Sec.~\ref{sec:Ehrenfest}. Therefore, here we conjecture that both frequencies of the dipole modes can be obtained with the stationary condition 
\begin{align}
\frac{\partial\omega^2}{\partial\theta}=&
\left(A_{+}B_{-}-A_{-}B_{+}\right)\cos2\theta 
\nonumber\\
&- 2\gamma_{12}\left(A_{-}+A_{+}\sin2\theta\right)=0,
\label{eq:stationarity}
\end{align}
whose solution will be denoted as $\omega_\pm$ (with mixing angles $\theta_\pm$).
Notice that in the non-interacting limit ($g_{12}=0$) we get $\omega_{-}=\omega_1$ ($\theta={\pi}/{4}$) and $\omega_{+}=\omega_2$ ($\theta={3\pi}/{4}$), which fixes the hierarchy of the two frequencies. 
Therefore, it is convenient to set $\omega_{1}\le\omega_{2}$ in order to identify $\omega_{-}$ and $\omega_{+}$ with the frequencies of low-lying and high-lying modes, respectively. 

\subsection{Gross-Pitaevskii equation and the Ehrenfest theorem}
\label{sec:Ehrenfest}

Let us now contextualize the above results within the framework of the Gross-Pitaevskii (GP) theory, namely the case in which the binary mixture is composed of two Bose-Einstein condensates. Their trapped evolution is described by a set of (three-dimensional) GP equations of the form
\begin{equation}
i\hbar\partial_{t}\psi_{i}(\bm{r},t)=\left[-\frac{\hbar^{2}}{2m_{i}}\nabla^{2}
+V^{tot}_{i}(\bm{r},t)\right]\psi_{i}(\bm{r},t)
\label{eq:GP}
\end{equation}
where $V^{tot}_i$ represents the total mean-field potential acting on the species $i$ ($\neq j$),
\begin{equation}
V^{tot}_{i}(\bm{r},t)= 
V_{i}^{ho}(\bm{r})
+ g_{ii}n_{i}(\bm{r},t) + g_{ij}n_{j}(\bm{r},t),
\end{equation}
with $n_{i}(\bm{r},t)\equiv|\psi_i(\bm{r},t)|^2$ and $\int n_{i}d^3r=N_{i}$.
For simplicity here we consider the harmonic potentials acting on the two species to be concentric. 

It is interesting to note that the sum-rule approach discussed in the previous section is equivalent to the prediction provided by the Ehrenfest theorem, namely,
\begin{equation}
\frac{d^{2}}{dt^2}\langle x\rangle_i=-\frac{1}{m_i}\langle \partial_{x}V^{tot}_i\rangle.
\end{equation}
This goes as follows. By defining $\delta x_{i}(t)\equiv \langle x\rangle_{i}(t)$ it is straightforward to get \footnote{We have used the fact that $\int n_i\partial_{x}n_id^3r=(1/2)\int d(n_i^{2})d^3r\equiv0$ because of the vanishing boundary conditions.}
\begin{equation}
 \ddot {\delta x_{i}}(t) = - \omega^{2}_{i}\delta x_{i}(t) -\frac{g_{ij}}{m_{i}N_{i}}\int n_{i}\partial_{x}n_{j}d^3r.
 \label{eq:ehrenfest}
\end{equation}
The integral in the last term can be evaluated explicitly by assuming that the densities translate rigidly, $n_{i}(x,t)= n_{0i}[x-\delta x_{i}(t)]$ (for ease of notation we omit the dependence on the ``transverse'' coordinates $y$ and $z$). This assumption is consistent with the sum-rule approach in the sense that it relies on the information encoded only in the ground state. Then, in the limit of small oscillations, we have
\begin{equation}
 n_{i}(x,t)\simeq n_{0i}(x) -\partial_{x}n_{0i}(x) \delta x_{i}(t).
\end{equation}
Within these approximations, by defining $\eta_{i}\equiv g_{12}I/(m_{i}N_{i})$, Eq.~(\ref{eq:ehrenfest}) can be recast in the following form \footnote{A related approach, based on the Boltzmann equation for the distribution functions of a Bose-Fermi mixture in the normal phase, can be found in Ref. \cite{asano2020}.}
\begin{equation}
 \ddot {\delta x_{i}} = - \left(\omega^{2}_{i} 
 - \eta_{i}\right)\delta x_{i} - \eta_{i}\delta x_{j},
\end{equation}
or, equivalently, in matrix form as
\begin{equation}
\ddot{\delta x}= -M\delta x,\quad
\delta x\equiv\begin{pmatrix}
 \delta x_{1} \\
 \delta x_{2}
 \end{pmatrix}
 \label{eq:eqofmotion}
\end{equation}
and
\begin{equation}
M=\begin{pmatrix}
 \omega_{1}^{2} - \eta_{1} & \eta_{1}
 \\
 \eta_{2} & \omega_{2}^{2} - \eta_{2}
 \end{pmatrix}.
\end{equation}

The frequencies $\omega_{\pm}$ of the two dipole modes 
are obtained as solutions of
\begin{equation}
 \det\left( M-\omega^2 I_2 \right)=0,
 \label{eq:det0}
\end{equation}
namely 
\begin{widetext}
\begin{equation}
\label{eq:modefreq}
 \omega^{2}_{\pm} = \frac12\sum_i(\omega^{2}_{i} - \eta_{i}) 
 \pm \frac12\sqrt{\left[\sum_i(\omega^{2}_{i} - \eta_{i})\right]^{2}
 + 4 \left(\omega_{1}^{2}\eta_{2} + \omega_{2}^{2}\eta_{1}\right) -4\omega_{1}^{2}\omega_{2}^{2}
 }.
\end{equation}
\end{widetext}
At this point we are able to prove the equivalence of the sum-rule approach in Sec.~\ref{sec:sumrules} and the Ehrenfest approach discussed above. Indeed, by plugging Eq.~(\ref{eq:omega_alpha}) into Eq.~(\ref{eq:det0}), it is not difficult to verify that the resulting equation corresponds to Eq.~(\ref{eq:stationarity}) of the sum-rule approach. Namely, the solutions of the latter equation are also solutions of Eq.~(\ref{eq:det0}), as well as those in Eq.~(\ref{eq:modefreq}), and therefore the two must coincide. Owing to this, all the results that we shall discuss in the following apply to both the Ehrenfest and sum-rule approaches. As a matter of fact, in the case of the dipole mode, the two approaches are completely equivalent.

\begin{figure*}
\centerline{\includegraphics[width=0.9\linewidth]{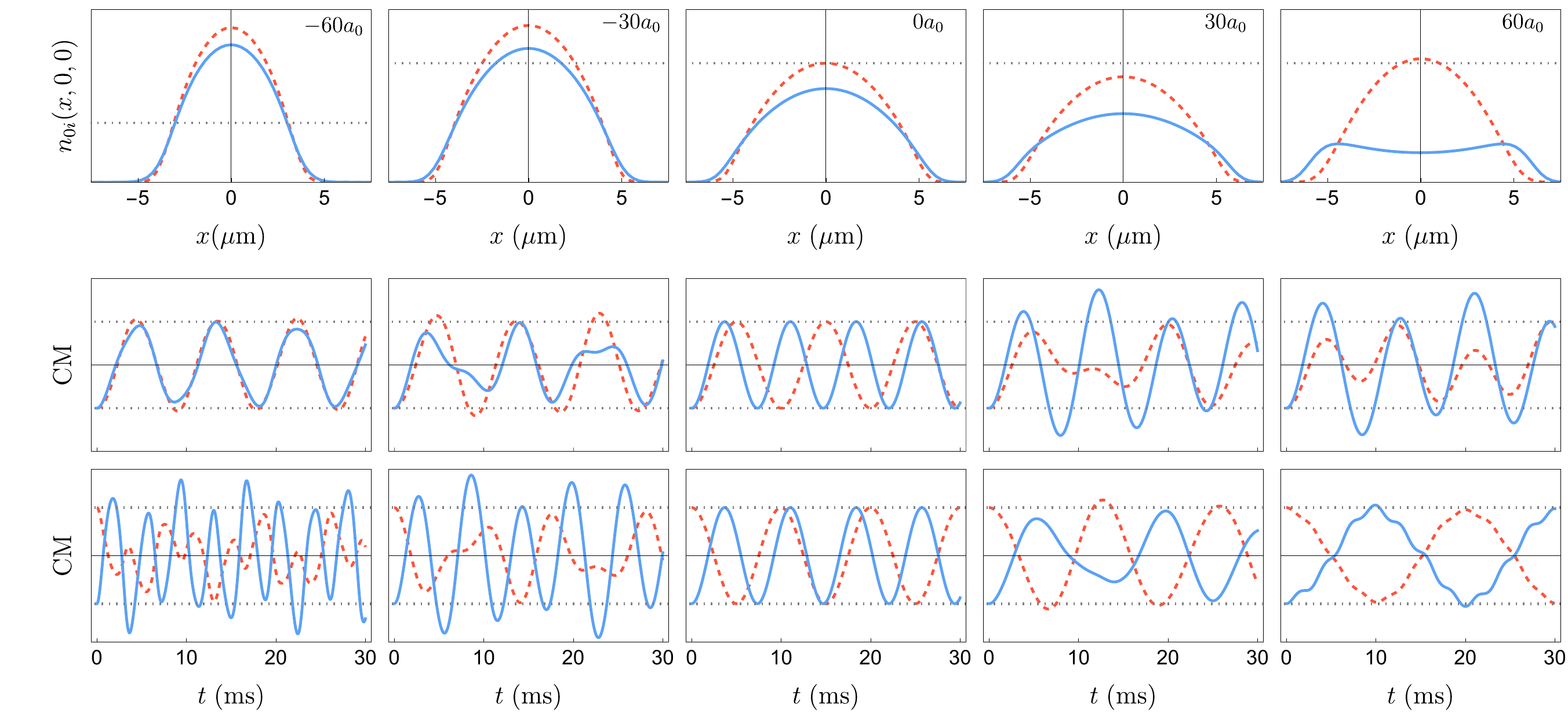}}
\caption{Top panel: Ground-state linear density profile $n_{0i}(x,0,0)$ (in arbitrary units) for the two components ($|1\rangle$: red dashed line, $|2\rangle$: blue solid line) as a function of the intra-species interaction $a_{12}$. As a reference we plot the dotted line corresponding to the maximum density of $|1\rangle$ for $a_{12}=0$. 
Middle and bottom: Center-of-mass (CM) evolution of the two condensates performing dipole oscillations, for the same values of $a_{12}$ as indicated in the top panels. The dotted horizontal lines correspond to the initial shift of each component (see text), which we consider either in the same direction (middle panel) or in opposite directions (bottom panel).}
\label{fig1}
\end{figure*}

In order to analyze the amplitude of the two modes contributing to the dipole oscillation of the mixture, let us now consider the solutions of Eq.~(\ref{eq:eqofmotion}).
First of all, we recall that the components of the eigenvectors of the matrix $M$, which we shall denote as $v_1=(v_{11},v_{12})$ and $v_2=(v_{21},v_{22})$, define the transformation matrix $P$ that diagonalizes $M$, $P^{-1}MP=\textrm{diag}(\omega_-^2,\omega_+^2)$, as
\begin{equation}
P=\left[ \begin{array}{cc}
 v_{11} & v_{21} \\
v_{12} & v_{22}
\end{array} 
\right].
\end{equation}
In the following, we shall take the vectors $v_1$ and $v_2$ both with the norm equal to $1$. Notice that in general the matrix $M$ is not symmetric, so that the eigenvectors are not necessarily orthogonal.
Then, by left multiplying Eq.~(\ref{eq:eqofmotion}) by $P^{-1}$, and introducing the ``diagonal'' coordinates $\xi=P^{-1}\delta x$, we can write the corresponding solutions as [we restrict the analysis to the case of initial vanishing velocity, $\dot{\xi}(0)=0$]
\begin{equation}
\begin{cases}
\xi_1(t)=A\cos(\omega_- t),
\\
\xi_2(t)=B\cos(\omega_+ t),
\end{cases}
\end{equation}
yielding
\begin{equation}
\begin{cases}
 x_1(t)= v_{11}A\cos(\omega_- t)+v_{21}B\cos(\omega_+ t) 
 \\
 x_2(t)= v_{12}A\cos(\omega_- t)+v_{22}B\cos(\omega_+ t).
\end{cases}
\label{ampiezze}
\end{equation}
From the above equations, we can express the relative weights of each of the two eigenmodes in the center-of-mass motion of the two condensates in terms of the ratios $r_-=v_{12}/v_{11}$ and $r_+=v_{21}/v_{22}$, respectively. By construction, these quantities are vanishing in the noninteracting limit, $r_{\pm}=0/1$ for $g_{12}=0$. It is also worth noticing that they do not depend on the amplitude of the oscillations (namely, they are independent of $A$, and $B$). 
By explicitly calculating the eigenvectors of the matrix $M$, we can express the two ratios as
 \begin{align}
 \label{eq:ratios}
 &r_\pm^{\pm1}=
 \frac{1}{2}\left[ \left( 1-\beta\right)+\frac{\omega_1^2-\omega_2^2}{\eta_{2}} \right]
 \\
 &\quad\pm \sqrt{\frac{1}{4} \left( 1+\beta\right)^2 + \left( 1 -\beta\right)\frac{\omega_1^2-\omega_2^2}{2\eta_{2}}+ \left( \frac{\omega_1^2-\omega_2^2}{2\eta_{2}}\right)^2},
 \nonumber
 \end{align}
where $\beta\equiv m_2N_2/(m_1N_1)$ represents the ratio between the total masses of the two components.

\section{A tunable bosonic mixture}
\label{sec:results}

In order to compare these analytical predictions with GP simulations, we consider an experimentally feasible system. In particular, we choose a mixture of $^{87}$Rb (species $|1\rangle$) and $^{41}$K (species $|2\rangle$) both in the $F=1,m_F=1$ state, where Feshbach resonances can be used to tune the interspecies interactions \cite{Thalhammer2008,derrico2019}.
Then, we have $m_1=m_{Rb}$, $m_2=m_K$, $a_{11}=100.4 a_0$, and $a_{22}=62a_0$, with $a_0$ being the Bohr radius.
As for the values of interspecies scattering length $a_{12}$, here we restrict the analysis to the regime in which the mixture is miscible and stable against mean-field collapse, such that $g_{12}^2<g_{11}g_{22}$ \cite{riboli2002}. The corresponding critical value of the scattering length is $|a_{12}^{c}|\simeq73.6a_{0}$.
The mixture is trapped in a spherical optical potential where the frequencies are $\omega_1=2\pi \times 100$ Hz, $\omega_2=2\pi \times136$ Hz $\simeq \omega_1\sqrt{\gamma m_1/m_2}$ and $\gamma\simeq0.87$ correspond to the relative polarizability of the two components for a trapping laser wavelength $\lambda=1.064$ $\mu$m.

\subsection{Balanced case}
\label{sec:balanced}

Let us start by considering the case of a balanced mixture, with $N_1=N_2=5\times 10^4$ (the effect of an imbalance in the population of the two components will be discussed in the following section). 
The equilibrium (ground-state) density profiles of the two components are shown in the top panel of Fig.~\ref{fig1} for different intraspecies interactions, ranging from $a_{12}=-60a_0$ to $a_{12}=+60a_0$ \footnote{The ground state of the system is obtained by minimizing the GP energy functional, see e.g. Ref. \cite{dalfovo1999}, by means of a steepest-descent algorithm \cite{press2007}. The typical size of the numerical box is $18$ $\mu$m per side, each discretized in 64 points.}. The corresponding center-of-mass oscillations, obtained by solving the time-dependent GP equations (\ref{eq:GP}) \footnote{The time-dependent GP equation is solved by using a fast-Fourier-transform split-step algorithm; see, e.g., Ref. \cite{jackson1998}.} are shown in the middle and bottom panels. In particular, dipole oscillations are induced by shifting the centers of mass of the two components by the same amount either in the same direction (middle panel) or in opposite directions (bottom panel).
In all cases, the dynamics takes place in the configuration in which the two traps are concentric (to which the ground state calculations refer), and the initial shift is small enough to be in the linear regime of small amplitude oscillations.

\begin{figure}[t]
\centerline{\includegraphics[width= 0.95\columnwidth]{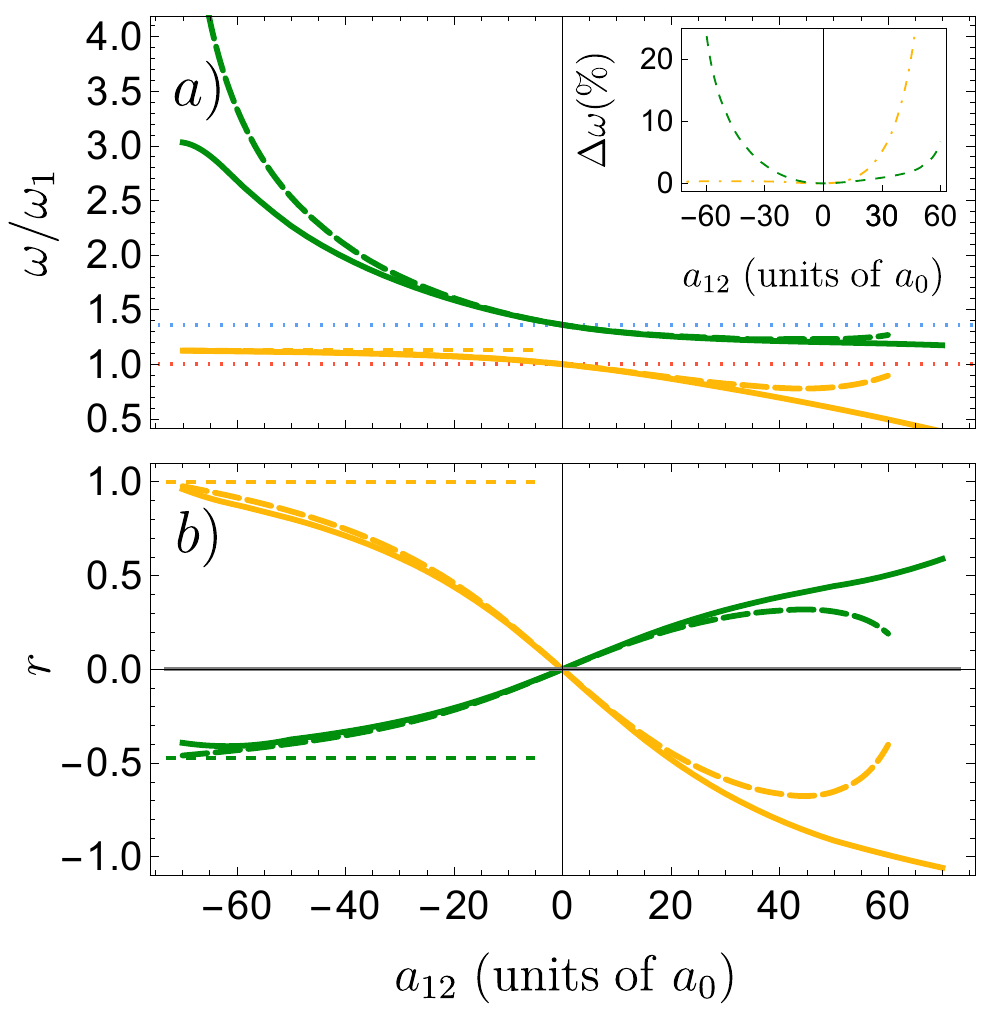}}
\caption{
Behavior of the high-lying and low-lying modes [in green (dark gray) and yellow (light gray), respectively] as a function of the interspecies scattering length $a_{12}$, for $|a_{12}|\leq |a_{12}^{c}|$. (a) The frequencies $\omega_{\pm}$ extracted from the GP simulations (solid lines) are compared to the prediction of the sum-rule and Ehrenfest approach in Eq.~(\ref{eq:modefreq}) (dashed lines). Their relative difference $\Delta\omega\equiv \left(\omega_{\pm}^{Ehr}-\omega_{\pm}^{GP}\right)/\omega_{\pm}^{GP}$ is shown in the inset. 
In the main panel we also show as a reference the trap frequencies $\omega_1$ (lower dotted red line) and $\omega_2$ (upper dotted blue line). Notice that all the frequencies are expressed in units of $\omega_1$.
(b) Relative weights of the two modes, $r_-$ [yellow (light gray)] and $r_+$ [green (dark gray)].
In both panels, the thin dashed lines represents the analytical limits of the Ehrenfest approach (see text).
} 
\label{fig2}
\end{figure}

Then, by fitting the center-of-mass dynamics with a double sinusoidal function, $f(t)=a\cos(\omega_- t) + b\cos(\omega_+ t)$, we extract the frequencies $\omega_\pm$ of the two modes and their relative amplitudes $r_\pm$ defined in the previous section. These quantities, obtained from the GP simulations, are plotted in Fig.~\ref{fig2} (solid lines) as a function of the interspecies scattering length $a_{12}$ \footnote{Notice that the behavior of the two dipole frequencies (as a function of $a_{12}$) presents the same general features as the case discussed in Ref. \cite{pu1998} for a Rb-Na mixture (see their Fig.~1).}, along with the predictions of the sum-rule and Ehrenfest approach (dashed lines).

As one can see from the inset in Fig.~\ref{fig2}a, the sum-rule and Ehrenfest approach provides an excellent result for the low-frequency mode, in the whole attractive regime, up to weakly repulsive interactions ($\simeq30a_0$).
It is worth noticing that in the limit of large attractive interactions [for $g_{12}<0$ and $|\eta_{i}|\gg \omega_i^2$, see Eq.~(\ref{eq:modefreq})], $\omega_-$ tends to the value
\begin{equation}
\omega_b \equiv \displaystyle\sqrt{\frac{N_1m_1\omega_1^2+N_2m_2\omega_2^2}{N_1m_1+N_2m_2}},
\label{eq:freq_limit_attractive}
\end{equation}
which corresponds to the center of mass frequency of the two condensates oscillating as a whole [thin dashed yellow (light gray) line in Fig.~\ref{fig2}(a)]. As for the high-frequency mode, 
the sum-rule and Ehrenfest approach well reproduces the GP results for $-30a_0\lesssim a_{12}\lesssim 50a_0$.

Instead, outside those regimes significant deviations are clearly visible. This has the following explanation. The sum-rule and Ehrenfest approach is based on only the knowledge of the ground-state density distribution, and therefore it cannot account for effects stemming from modifications of the condensate shape during the dynamics. This is particularly clear in the derivation based on the Ehrenfest theorem, which follows from the assumption that the density profiles of the two condensates perform only a rigid translation during their motion. Clearly, this condition is violated for out-of-phase oscillations [corresponding to negative values of $r$; see Fig.~\ref{fig2}(b)] in the regime where the two components interacts strongly (for both attractive and repulsive interactions). In addition, one should also consider that when the two components are trapped by different frequencies and interact with each other, as in the present case, the Kohn theorem \cite{kohn1961} does not hold. This implies not only that even the frequency of the in-phase mode becomes interaction-dependent but also that when the two components approach the phase-separation regime (for repulsive interactions), the sum-rule and Ehrenfest approach may fail again for the same reason as before. Indeed, significant modification of the density may be expected for sufficiently strong interactions, whenever the two components do not move as a whole.

Let us now turn to the discussion of the amplitudes of the two modes. In particular, we consider the relative weights $r_{\pm}$ defined in the previous section, which are shown in Fig.~\ref{fig2}(b) as a function of $a_{12}$.
Several comments are in order. First of all, it is evident that the character of the two modes is fixed by the sign of $a_{12}$. Namely, for attractive interactions the low-lying mode is associated with in-phase oscillations ($r_{-}>0$), whereas the high-lying mode determines an out-of-phase dynamics ($r_{+}<0$). On the repulsive side, the situation is reversed. 
Then, we notice that the Ehrenfest prediction for $r_{\pm}$ well reproduces the GP results in a wide range of interactions, for $a_{12}^c \lesssim a_{12}\lesssim 20a_{0}$. In addition, we can also derive an analytic estimate for the asymptotic values in the strongly attractive regime (for $a_{12}\to a_{12}^c$). Indeed, in the limit $\eta_{2}^{-1}(\omega_1^2-\omega_2^2)\ll 1-\beta$, from Eq.~(\ref{eq:ratios}) we obtain:
\begin{equation}
r_+=-\beta, 
\quad
r_-=1,
\end{equation}
which are shown as thin dashed lines in Fig. \ref{fig2}. Remarkably, the fact that the low-lying mode contributes with equal weight to the center-of-mass oscillation of the two condensates ($r_-=1$) is consistent with the result in Eq.~(\ref{eq:freq_limit_attractive}), namely, the fact that the two condensates oscillate as a whole.

\subsection{Effect of the population imbalance}
\begin{figure}[t]
\centering
\includegraphics[width=0.95\columnwidth]{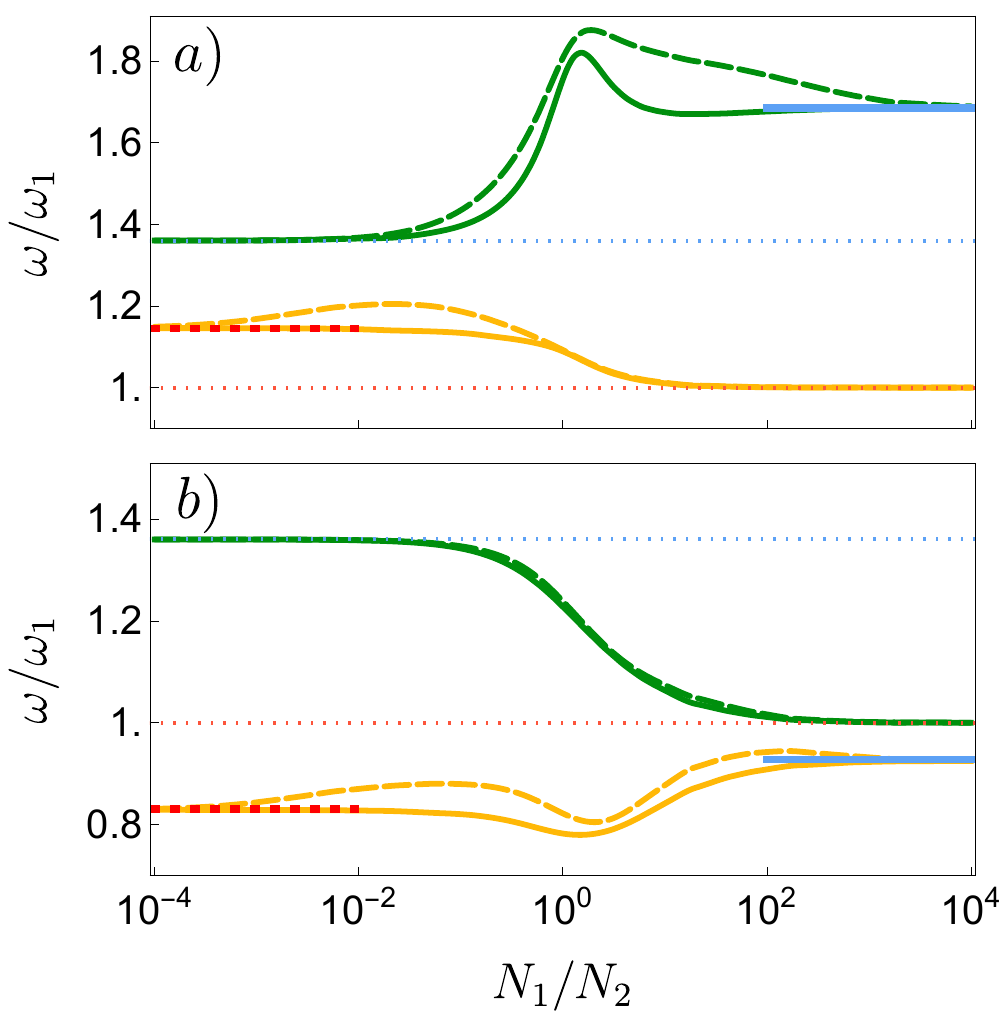}
\caption{
Behavior of the dipole frequencies $\omega_{\pm}$ [in green (dark gray) and yellow (light gray), respectively] obtained from the GP simulations (solid lines) and from the sum-rule and Ehrenfest approach
(dashed lines), as a function of $N_1/N_2$ (in log scale, with $N_1+N_2=10^5$), for (a) $a_{12}=-30 a_0$  and (b) $a_{12}=+30 a_0$.
For comparison we also show the two trapping frequencies $\omega_i$ as dotted horizontal lines, and the two limiting values $\tilde{\omega}_{i}$ for the minority component in Eq.~(\ref{eq:limitfreq}) as horizontal lines ($|1\rangle$: red dashed line, $|2\rangle$: blue solid). All the frequencies are expressed in units of $\omega_1$.}
\label{fig3}
\end{figure}

We now turn to a discussion of the role of the population imbalance of the two components. To this end, as illustrative cases, we consider a weakly attractive mixture at $a_{12}=-30a_{0}$, and the corresponding repulsive case at $a_{12}=+30a_{0}$ (as in Fig.~\ref{fig1} for the balanced mixture). We keep the total number fixed to $N_1+N_2=10^5$, as in Sec.~\ref{sec:balanced}, and we vary the ratio $N_1/N_2$ in the range $[10^{-4},10^4]$. 
In analogy with the previous discussion, in Fig.~\ref{fig3} we compare the frequencies $\omega_\pm$ of the two modes computed from the GP simulations with the predictions of the sum-rule and Ehrenfest approach, this time as a function of $N_1/N_2$.

Let us first discuss the attractive case at $a_{12}=-30 a_0$, in Fig.~\ref{fig3}(a). The plot clearly shows that when the mixture becomes strongly imbalanced, $N_1/N_2\ll1$ or $N_1/N_2\gg1$, the frequencies of the two eigenmodes match the expected values for the \textit{impurity limit} discussed in Refs. \cite{sartori2013,wilson2021}: The majority component oscillates at its own (bare) trap frequency, whereas the frequency of the minority component $i$ is shifted to the following value:
\begin{equation}
 \tilde{\omega}_{i}=\omega_{i}\sqrt{1-\frac{g_{12}m_j\omega_j^2}{g_{jj}m_i\omega_i^2}} \qquad(i\neq j).
 \label{eq:limitfreq}
\end{equation}
The latter expression, which is valid when the majority component $j$ can be accurately described within the Thomas-Fermi (TF) approximation, can be obtained either from the GP equation -- by considering the effective harmonic potential acting on the minority component due to the mean-field interaction with the other species (neglecting the back action of the former on the latter) \cite{sartori2013,wilson2021} -- or, equivalently, from the sum-rule and Ehrenfest approach. Indeed, in the TF limit for the majority species, the integral $I$ entering Eq.~(\ref{eq:omega_alpha}) and the definition of $\eta_i$ can be expressed as
\begin{equation}
I=-\frac{m_{j}\omega_{j}^{2}}{g_{jj}}\int x\partial_{x}n_{0i}d^3r
=\frac{m_{j}\omega_{j}^{2}N_{i}}{g_{jj}},
\label{eq:int}
\end{equation}
so that one can easily recover the two asymptotic values from Eq.~(\ref{eq:modefreq}), by taking the limit $N_i/N_j\ll1$.
In this regard, it is worth noticing (see Fig.~\ref{fig3}) that the sum-rule and Ehrenfest prediction reaches those values much later than the GP case. Indeed, a close inspection reveals that the assumption of neglecting the effect of the interspecies interactions on the density profile of the majority component is justified only for $N_i/N_j\lesssim10^{-3}$, where the integral $I$ gets close to the noninteracting value [notice that the expression in Eq.~(\ref{eq:int}) does not depend on $g_{12}$]. 
As a matter of fact, the minority component is still able to produce local density modifications on the majority component (which, in turn, produce a back-action on the former) in the regime of intermediate imbalances ($N_i/N_j\sim10^{-2}$), where its effect on the dipole motion of the majority component can already be safely neglected.
This suggests that the GP results converge earlier to the asymptotic limit of Eq.~(\ref{eq:limitfreq}) as a consequence of additional contributions to the mean-field renormalization of the harmonic potential, beyond the simple ansatz discussed above.

Figure~\ref{fig3}(a) also shows that, for attractive interactions, the low-lying mode is mostly dominated by species $|1\rangle$ (whose characteristic frequency is represented by the horizontal red dashed line) and the high-lying one by $|2\rangle$ (horizontal blue solid line).
For the low-lying mode, significant deviations between the GP result and the sum-rule and Ehrenfest prediction (up to $\sim10\%$) are visible in the regime with a slight minority of the dominant species ($|1\rangle$), whereas for the high-lying mode they are present in the whole intermediate regime (although at different degrees). 
The reason for this behavior possibly lies in the fact that when the two components have different spatial extensions the system is characterized by different regions where the two clouds are superimposed or not, further complicating their coupled dipole dynamics, as suggested for the case of Bose-Fermi mixtures \cite{maruyama2008}. 
As discussed in the previous section, this effect is amplified in the case of out-of-phase oscillations, namely, for the high-lying mode (at $a_{12}=-30 a_0$).

As for the repulsive case at $a_{12}=+30 a_0$, in Fig.~\ref{fig3}(b), the most relevant difference from the previous case is the fact that the high-lying mode is always associated with the majority species, and the low-lying mode is associated with the minority one. 
This is signaled by the two asymptotic values of each frequency: The limits of $\omega_+$ are the bare trap frequencies of the two species, whereas $\omega_-$ tends to the values fixed by Eq.~(\ref{eq:limitfreq}) (for $i=1,2$).
This behavior also provides an explanation of the different degrees of accuracy of the sum-rule and Ehrenfest approach, in this interaction regime. 
Indeed, the high-lying mode, being associated with the majority component (less affected by the presence of the other) and also having an in-phase character (see Fig.~\ref{fig2}), is very well reproduced. Conversely, significant deviations are appreciable for the low-lying frequency, for the same reasons mentioned above for the corresponding case in Fig.~\ref{fig3}(a).

\section{Conclusions}

We have presented a systematic discussion of the sum-rule approach for the dipole collective modes of a harmonically trapped bosonic degenerate mixture. We have shown that the sum-rule prediction for the dipole frequency can be alternatively obtained from a mean-field effective approach based of the Gross-Pitaevskii (GP) equation and the Ehrenfest theorem. Remarkably, the latter also provides an estimate for the amplitude of the collective modes. Then, by direct comparison with full GP calculations, we have analyzed the accuracy of the sum-rule and Ehrenfest predictions as a function of the interspecies interaction, in the whole regime in which the mixture is miscible and stable against mean-field collapse.

For a balanced mixture, namely with the same population in each species, we have found that the sum-rule and Ehrenfest approach is quite accurate for not too strong interspecies interactions, whereas some deviation from the exact GP result becomes significant for the low-lying dipole mode in the repulsive regime, especially for the high-lying mode for strongly attractive interspecies interactions. This behavior is associated with the out-of-phase character of the two oscillation modes, and it can be explained in terms of a breaking of the self-similarity assumption in the dynamical evolution of the density distributions, on which the effective Ehrenfest approach is based.

We have also analyzed the effect of a population imbalance, for both moderately attractive and repulsive mixtures, finding that the sum-rule and Ehrenfest approach well reproduces the frequency of the modes dominated by the majority component, and it is especially accurate in reproducing the high-lying frequency in the repulsive case. 
Explanations for the discrepancies observed in other regimes have been provided, along with a comparison with the analytic prediction in the limit of strongly imbalance mixtures.

Our analysis provides a framework for further theoretical studies aimed at benchmarking the accuracy of sum-rule or related approaches in multi-component systems, and it may prove valuable for future experimental studies of dipole excitations in bosonic mixtures.

\begin{acknowledgments}
We thank A.~Burchianti, L.~Cavicchioli, and F.~Minardi for useful comments and suggestions.
MM acknowledges support from the  Grant PGC2018-101355-B-I00 funded by MCIN/AEI/10.13039/501100011033 and by ``ERDF A way of making Europe'', and from the Basque Government through Grant No. IT1470-22.
\end{acknowledgments}

\bibliography{sumrules}

\end{document}